\documentclass{aa}

\usepackage{graphicx}
\usepackage{txfonts}

\usepackage{amsmath}    % Advanced maths commands
\usepackage{amssymb}    % Extra maths symbols
\usepackage{mathtools}
\usepackage{multicol} 
\usepackage{multirow} 
\usepackage{threeparttable}
\usepackage{longtable}
\usepackage{tabto}
\usepackage{adjustbox}
\usepackage{color}
\usepackage{natbib,twoopt}
\usepackage[hyphenbreaks]{breakurl}
\usepackage[breaklinks]{hyperref}      %% to avoid \citeads line fills, add "draft" 
\bibpunct{(}{)}{;}{a}{}{,}             %% natbib format for A&A and ApJ
\definecolor{cobalt}{rgb}{0.06, 0.2, 0.65}
\hypersetup{
  colorlinks,
  citecolor=blue,
  linkcolor=blue,
  urlcolor=blue,
}
\makeatletter
  \newcommandtwoopt{\citeads}[3][][]{\href{http://adsabs.harvard.edu/abs/#3}%
    {\def\hyper@linkstart##1##2{}%
     \let\hyper@linkend\@empty\citealp[#1][#2]{#3}}}
  \newcommandtwoopt{\citepads}[3][][]{\href{http://adsabs.harvard.edu/abs/#3}%
    {\def\hyper@linkstart##1##2{}%
     \let\hyper@linkend\@empty\citep[#1][#2]{#3}}}
  \newcommandtwoopt{\citetads}[3][][]{\href{http://adsabs.harvard.edu/abs/#3}%
    {\def\hyper@linkstart##1##2{}%
     \let\hyper@linkend\@empty\citet[#1][#2]{#3}}}
  \newcommandtwoopt{\citeyearads}[3][][]%
    {\href{http://adsabs.harvard.edu/abs/#3}
    {\def\hyper@linkstart##1##2{}%
     \let\hyper@linkend\@empty\citeyear[#1][#2]{#3}}}
\makeatother

\newcommand{\Msun}{M$_{\odot}$}
\newcommand{\Rsun}{R$_{\odot}$}
\newcommand{\Msunyr}{M$_{\odot}$~yr$^{-1}$}

\definecolor{smalt(darkpowderblue)}{rgb}{0.0, 0.2, 0.6}
\definecolor{forestgreen(traditional)}{rgb}{0.0, 0.5, 0.0}
\defcitealias{RVJ}{RVJ}

\defcitealias{MD17}{MD17}

\newcommand{\hf}{helium shell flash}
\newcommand{\hfs}{helium shell flashes}

\newcommand{\koi}{KOI~3278}

\begin{document}

   \title{Formation of long-period post-common-envelope binaries}

   \subtitle{II. Explaining the self-lensing binary KOI 3278}

   \titlerunning{Formation of long-period PCEBs II. Explaining the self-lensing binary KOI 3278}

   \author{Diogo Belloni\inst{1}
           \and
           Matthias R. Schreiber\inst{1,2}
           \and
           Monica Zorotovic\inst{3}
          }

    \authorrunning{Belloni, Schreiber \& Zorotovic}

   \institute{Departamento de F\'isica, Universidad T\'ecnica Federico Santa Mar\'ia, Av. España 1680, Valpara\'iso, Chile\\
              \email{diogobellonizorzi@gmail.com}
              \and
              Millenium Nucleus for Planet Formation, Valpara{\'i}so, Chile
              \and
              Instituto de F\'isica y Astronom\'ia, Universidad de Valpara\'iso, Av. Gran Breta\~na 1111, Valpara\'iso, Chile
             }

   \date{Received...; accepted ...}

% \abstract{}{}{}{}{} 
% 5 {} token are mandatory
 
  \abstract
   % context heading (optional), leave it empty if necessary  
   {
   The vast majority of close binaries containing a compact object, including the progenitors of supernovae\,Ia and at least a substantial fraction of all accreting black holes in the Galaxy, form through common-envelope (CE) evolution. Despite this importance, we struggle to even understand the energy budget of CE evolution. For decades, observed long-period post-CE binaries have been interpreted as evidence of additional energies contributing during CE evolution. We have recently shown that this argument is based on simplified assumptions for all long-period post-CE binaries containing massive white dwarfs (WDs). The only remaining post-CE binary star that has been claimed to require contributions from additional energy sources to understand its formation is KOI~3278.
   }
   {
   Here, we address in detail the potential evolutionary history of KOI~3278. In particular, we investigate whether extra energy sources, such as recombination energy, are indeed required to explain its existence.
   }
   {
   We used the 1D stellar evolution code MESA to carry out binary evolution simulations and searched for potential formation pathways for KOI~3278 that are able to explain its observed properties.
   }
   {
   We find that KOI~3278 can be explained if the WD progenitor filled its Roche lobe during a helium shell flash. In this case, the orbital period of KOI~3278 can be reproduced if the CE binding energy is calculated taking into account gravitational energy and thermodynamic internal energy. While the CE evolution that led to the formation of KOI~3278 must have been efficient -- that is, most of the available orbital energy must have been used to unbind the CE -- recombination energy is not required. 
   }
   {
   We conclude that currently not a single observed post-CE binary requires one to assume 
   that energy sources other than gravitational and thermodynamic energy are contributing to CE evolution. KOI~3278, however, remains an intriguing post-CE binary as, unlike its siblings, understanding its existence requires highly efficient CE ejection.
   }


   \maketitle
%
%-------------------------------------------------------------------

%=============================
%=============================
%        BODY
%=============================
%=============================

%%%%%%%%%%%%%%%%
% NEW SECTION
%%%%%%%%%%%%%%%%
\section{Introduction}
\label{introduction}

A significant fraction of all compact binaries, which includes low-mass X-ray binaries and double degenerate objects, are thought to form through common-envelope (CE) evolution \citep[e.g.][]{Paczynski_1976,Ivanova_REVIEW,BelloniSchreiberChapter}.
Despite some recent progress, we still struggle to reliably predict the outcome of this fundamental evolutionary phase from hydrodynamic simulations.
Therefore, typically a simple energy equation, relating the binding energy of the envelope to the change in orbital energy scaled with an efficiency, $\alpha_{\mathrm{CE}}$, is used in binary population models. 
Observed post-CE binaries have frequently been used to derive constraints on the importance of different terms in this energy balance and on the CE efficiency.

For the vast majority of observed post-CE systems, with typical orbital periods of hours to a few days, it has been found that CE evolution with an efficiency of $\alpha_{\mathrm{CE}}\simeq0.3$, and only taking into account thermal and gravitational energies, convincingly explains the observations
\citep{Zorotovic_2010, Zorotovic_2022,hernandezetal22-1}. 
However, two types of post-CE binaries have challenged this picture; namely, 
the Kepler Object of Interest 3278 (\koi) \citep{Zorotovic_2014KOI} and long-period systems with oxygen-neon white dwarfs (WDs) \citep{Davis_2010,Zorotovic_2010,Yamaguchi_2024}.
These two types of systems have been claimed to require contributions from additional energy sources, such as recombination energy, during CE evolution to explain their current characteristics, in particular their long orbital periods.

We have recently shown that there is no need to invoke extra energy sources to explain the long orbital period of post-CE binaries with oxygen-neon WDs if CE evolution was triggered by dynamically unstable mass transfer from a highly evolved thermally pulsing asymptotic giant branch (TP-AGB) star \citep[][]{Belloni_2024b}.
In this case, at the onset of mass transfer, the mass of the envelope of the donor is comparable to the mass of its core, which means that the envelope is sufficiently loosely bound and can be ejected due to the input of relatively little orbital energy, resulting in a long-period post-CE binary with a massive WD.
The single remaining system that has been claimed to provide evidence for contributions from additional energy sources is thus \koi.

\koi~consists of a G-type main-sequence (MS) star eclipsed by a low-mass carbon-oxygen WD companion.
It was classified as a candidate system for showing a planetary transit signal due to the repeated occultation of the WD as it passes behind the G-type MS star \citep{Burke2014,Tenenbaum2014}.
Data from the Quasiperiodic Automated Transit Search algorithm \citep{CarterAgol2013} revealed 16 occultations as well as 16 pulses of brightening occurring almost half an orbital period later with the same period and duration as the occultations.
\citet{KruseAgol_2014} interpreted these brightenings as the gravitational microlensing effect, which produces a magnification of the G-type star as a WD passes in front of it in a nearly circular orbit.
The most recent robust measurements of the parameters of \koi~were made by \citet{Yahalomi_2019}, who used a joint Einsteinian microlensing and Newtonian radial velocity model.

In this work, which is the second of a series of papers dedicated to long-period post-CE binaries, we investigate whether a formation pathway that includes detailed calculations of the TP-AGB phase can explain not only the existence of the long period post-CE binaries containing oxygen-neon WDs \citep{Belloni_2024b}, but also the properties of \koi, which contains a WD of a much lower mass, without additional energy sources.  
We find that for post-CE binaries with low-mass carbon-oxygen WDs ($\lesssim0.55$~\Msun) and sufficiently long orbital periods, like \koi, the onset of CE evolution has to occur during a \hf, when the WD progenitor is at the beginning of the TP-AGB phase, to explain their current orbital periods without considering energy sources in addition to orbital and thermodynamic internal energy.

%%%%%%%%%%%%%%%%%%%%%%%%%%%%%%%%
%%%%%%%%%%%%%%%%%%%%%%%%%%%%%%%%
% NEW SECTION
%%%%%%%%%%%%%%%%%%%%%%%%%%%%%%%%
%%%%%%%%%%%%%%%%%%%%%%%%%%%%%%%%
\section{Model assumptions}
\label{Method}

We used the MESA code \citep{Paxton2011,Paxton2013,Paxton2015,Paxton2018,Paxton2019,Jermyn2023}, version r15140, to simulate binary evolution.\footnote{The files \texttt{run\_star\_extras.f90} and \texttt{run\_binary\_extras.f90} as well as the inlists needed for the simulations are available at \href{https://zenodo.org/records/10841636}{https://zenodo.org/records/10841636}}
We adopted the grey Eddington T(tau) relation to calculate the outer boundary conditions of the atmosphere, using a uniform opacity that is iterated to be consistent with the final surface temperature and pressure at the base of the atmosphere.
We assumed an auto-extended scheme for the nuclear network, which automatically extends the net as needed.
We included mass loss through stellar winds, adopting the \citet{Reimers_1975} prescription during pre-AGB evolution with a wind efficiency parameter equal to $0.5$, and the prescription proposed by \citet{VW93} for wind mass-loss during AGB evolution.
Although we set the rotation of the zero-age MS stars to zero, we allowed it to vary during binary evolution according to the MESA standard prescription.
For the stability criterion of convection, we adopted that proposed by \citet{Ledoux_1947}.
We treated convective regions using the scheme by \citet{Henyey_1965} for the mixing-length theory.
We used the predictive mixing scheme with the parameters suggested by \citet{Ostrowski_2021}.
We also included rotationally and non-rotationally-induced mixing, namely angular momentum mixing, Solberg-Hoiland, secular shear instability, Eddington-Sweet circulation, Goldreich-Schubert-Fricke, and Spruit-Tayler dynamo \citep{Heger_2000,Heger_2005}.
We allowed changes in the angular velocities of the stars due to tidal interactions.
The Roche lobe radius of each star was computed using the fit of \citet{Eggleton1983}.
The mass transfer rates due to Roche lobe overflow were determined following the prescription of \citet{Ritter1988}.
We adopted the Bondi-Hoyle-Lyttleton prescription \citep{Hoyle_1939,Bondi_1944} for wind accretion.
All other MESA parameters not mentioned here were fixed and set to their default values in version r15140.

For CE evolution, we adopted the so-called energy formalism and the relation put forward by \citet{Iben_Livio_1993}, in which the outcome of CE evolution is approximated by the balance between the change in the orbital energy and the envelope binding energy, given by

\begin{equation}
E_{\rm bind} \ = \ 
\alpha_{\rm CE} \ \Delta E_{\rm orb} \ = \ - \ 
\alpha_{\rm CE} \
 \left( \, \frac{G\,M_{\rm d,c}\,M_{\rm a}}{2\,a_f}  \ - \ 
           \frac{G\,M_{\rm d,c}\,M_{\rm a}}{2\,a_i} \, \right) \ ,
\label{Eq:AlphaCE}
\end{equation}

\noindent
where $E_{\rm bind}$ is the envelope binding energy, $E_{\rm orb}$ is the orbital energy, $G$ is the gravitational constant, $M_{\rm d,c}$ is the mass of the hydrogen-free core of the donor, $a_i$ is the semimajor axis at the onset of the CE evolution, $a_f$ is the semimajor axis after CE ejection, and $\alpha_{\rm CE}$ is a parameter corresponding to the fraction of the difference in orbital energy (before and after CE evolution) that unbinds the envelope.

We calculated the binding energy by integrating over the star envelope from the helium core boundary (i.e., the radius at which the helium mass fraction is 0.1) to the surface of the star; that is,

\begin{equation}
E_{\rm bind} \ = \ 
- \int_{M_{\rm d,c}}^{M_{\rm d}} \frac{G \; m}{r(m)} \; {\rm d}m 
 \ + \  
\int_{M_{\rm d,c}}^{M_{\rm d}} \varepsilon_{\rm int}(m) \; {\rm d}m \ ,
\label{Eq:AlphaINT}
\end{equation}

\noindent
where $m$ is the mass coordinate, $r$ is the radius at a given mass coordinate, and $\varepsilon_{\rm int}$ is the thermodynamic internal energy per unit mass, which is related to the equation of state and contains terms such as the thermal energy of the gas and the radiation energy, but not the recombination energies.\footnote{We computed the thermodynamic internal energy following closely \citet{Hirai_2022}, who made their MESA files available in \href{https://zenodo.org/records/7066430}{https://zenodo.org/records/7066430}.}
We assumed that the onset of the CE evolution takes place when the mass transfer rate exceeds $10^{-2}$~\Msunyr.

\begin{table*}
\centering
\caption{Evolution of a zero-age MS binary toward \koi. For the pre-CE evolution, we used the MESA code with the assumptions described in Sect.~\ref{Method}. For the CE evolution, we computed the post-CE orbital period using Eqs.~\ref{Eq:AlphaCE} and \ref{Eq:AlphaINT}, assuming ${\alpha_{\rm CE}\approx0.97}$. The terms $M_1$ and Type$_1$ are the mass and stellar type$^{a}$ of the WD progenitor, respectively, while $M_2$, $R_2$, $T_{\rm eff,2}$, $\log{}g_2$, and Type$_2$ are the mass, radius, effective temperature, log-scaled gravity, and stellar type of its companion, respectively. $P_{\rm orb}$ is the orbital period and the last column corresponds to the event occurring in the binary at the given time in the first column. The row in which the binary has the present-day properties of \koi~is highlighted in boldface.}
\label{Tab:FormationChannel}
\setlength\tabcolsep{7pt} % default value: 6pt
\renewcommand{\arraystretch}{1.25} % Default value: 1
\begin{tabular}{r c c c c c r c r l}
\hline
\noalign{\smallskip}
 Time  &   $M_1$    &   $M_2$    &    $R_2$  &   $T_{{\rm eff},2}$    &    $\log{}g_2$  & Type$_1$  & Type$_2$ & $P_{\rm orb}$ & Event\\
 (Myr) & (M$_\odot$)&(M$_\odot$) &(R$_\odot$)&    (K)                 &       (cm\,s$^{-2}$)          &           &          &  (d)        &      \\
\hline
\noalign{\smallskip}
     0.0000  &  1.500  &  0.900  &  0.790    &   5218  &  4.597    & MS     & MS  &    920.0000  &  zero-age MS binary \\
  2170.1105  &  1.497  &  0.900  &  0.817    &   5257  &  4.568    & SG     & MS  &    922.2610  &  change in primary type \\
  2524.7782  &  1.496  &  0.900  &  0.821    &   5267  &  4.564    & FGB    & MS  &    923.0199  &  change in primary type \\
  2893.2300  &  1.358  &  0.909  &  0.841    &   5326  &  4.547    & CHeB   & MS  &    953.7384  &  change in primary type \\
  3017.6612  &  1.333  &  0.910  &  0.835    &   5319  &  4.554    & E-AGB  & MS  &    978.3332  &  change in primary type \\
  3029.7819  &  1.333  &  0.910  &  0.835    &   5319  &  4.554    & TP-AGB & MS  &    989.6612  &  change in primary type \\
  3029.8615  &  1.184  &  0.910  &  0.835    &   5319  &  4.553    & TP-AGB & MS  &    978.1196  &  begin RLOF \\
  3029.8615  &  1.184  &  0.910  &  0.835    &   5319  &  4.553    & TP-AGB & MS  &    978.1196  &  onset of CE evolution \\
  3029.8615  &  0.531  &  0.910  &  0.835    &   5319  &  4.553    & WD     & MS  &     88.1540  &  end RLOF \\
    \textbf{5219.1031}  &  \textbf{0.531}  &  \textbf{0.910}  &  \textbf{0.861}    &  \textbf{5385}  &  \textbf{4.527}    & \textbf{WD}     & \textbf{MS} &   \textbf{88.1805}  &  \textbf{binary looks like \koi} \\
\noalign{\smallskip}
\hline
\end{tabular}
\tablefoot{\tablefoottext{a}{Abbreviations:
MS~(main~sequence~star),
SG~(subgiant~star),
FGB~(first giant branch star),
CHeB~(core helium burning),
E-AGB~(early asymptotic giant branch star),
TP-AGB~(thermally pulsing~asymptotic~giant~branch~star),
WD~(white~dwarf),
RLOF~(Roche~lobe~overflow),
CE~(common~envelope).}}
\end{table*}

%%%%%%%%%%%%%%%%%%%%%%%%%%%%%%%%
%%%%%%%%%%%%%%%%%%%%%%%%%%%%%%%%
% NEW SECTION
%%%%%%%%%%%%%%%%%%%%%%%%%%%%%%%%
%%%%%%%%%%%%%%%%%%%%%%%%%%%%%%%%
\section{Numerical procedure}

Our search toward an evolutionary model of \koi~can be divided into three steps: (i) we varied the metallicity and mixing length of single-star models to find a fit for the secondary star; (ii) we ran a grid of binary models to identify a region in the parameter space that could offer a solution; and (iii) we used a finer grid to find an evolutionary track that produces a binary resembling \koi.

%::::::::::::::::::::::::::::::
% Main Sequence Star
%::::::::::::::::::::::::::::::

For the first step, we used the properties of the MS star.
We ran a grid of single-star models, varying the metallicity and the mixing length, as these two parameters are the most important in determining the stellar effective temperature, radius, and log$g$.
For the mixing length, in units of the local pressure scale height, ${H_{\rm p}}$, we adopted values from $1.7$ to $2.3$, in steps of $0.1$.
We adopted metallicities ranging from $0.020$ to $0.030$, in steps of $0.001$.
We assumed an initial mass of $0.912$~\Msun for the MS star, which is close to the observed value because the secondary is expected to accrete only a negligible amount of mass due to either wind accretion or atmospheric Roche lobe overflow before the WD progenitor fills its Roche lobe.
We found that for a metallicity comparable to the observed one (i.e., ${Z\approx0.024}$), the mixing length has to be ${\approx2.0\,H_{\rm p}}$.

Having determined the metallicity and the mixing length, we searched for a binary model able to explain the present-day properties of \koi~by running first a broad grid of models and subsequently finer grids to properly cover the parts of the parameter space that might offer solutions for \koi.
In all of the simulations, we initially assumed a binary consisting of two zero-age MS stars in a circular orbit.
We fixed the initial mass of the WD progenitor to $1.5$~\Msun, which is a value high enough to avoid dynamically stable mass transfer when it fills its Roche lobe and 
%at the same time is 
low enough for the core mass to not exceed the observed WD mass.
We varied the mass of the zero-age MS companion around $0.911$~\Msun~and the zero-age orbital period around $1000$~d.
%, the metallicity, the mixing length, the amount of core overshooting, the efficiency of semiconvective, predictive and thermohaline mixing, the criterion to determine whether or not a region inside the star is stable against convection, the inclusion or not of gravitational settling and chemical diffusion.
%
The metallicity and the mixing length were set to the values we found in the previous step; that is, $0.024$ and $2.00$, respectively.
%
%All other stellar and binary evolution parameters are fixed and set to their default values.
%
%For the mixing length, in units of the local pressure scale height ${H_{\rm p}}$, we adopted $1.8$, $2.0$ and $2.2$.
%
%For the core overshooting, we adopted for the extent of the overshoot region $0.0$ (i.e., no overshooting) and ${0.016~H_{\rm p}}$.
%
%For the stability criterion of Ledoux [FINISH].
%
%When the Ledoux criterion is adopted, we adopted for the 
%

%WD progenitor fills its Roche lobe.
%
%All other MESA parameters not mentioned here
%and in the Appendix~\ref{MESA} 
%are fixed and set to their default values of the version r15140.

%%%%%%%%%%%%%%%%%%%%%%%%%%%%%%%%
%%%%%%%%%%%%%%%%%%%%%%%%%%%%%%%%
% NEW SECTION
%%%%%%%%%%%%%%%%%%%%%%%%%%%%%%%%
%%%%%%%%%%%%%%%%%%%%%%%%%%%%%%%%
%\section{CE evolution}
%\label{CE}

%%%%%%%%%%%%%%%%%%%%%%%%%%%%%%%%
%%%%%%%%%%%%%%%%%%%%%%%%%%%%%%%%
% NEW SECTION
%%%%%%%%%%%%%%%%%%%%%%%%%%%%%%%%
%%%%%%%%%%%%%%%%%%%%%%%%%%%%%%%%
\section{Potential formation pathway for KOI~3278}
\label{KOI}

In what follows, we discuss a formation pathway for \koi~that does not rely on the inclusion of recombination energy (or other additional energy sources) during CE evolution. 
The mass estimated by \citet{Yahalomi_2019} for the WD in \koi~is ${\approx0.53}$~\Msun, which is low for a carbon-oxygen WD.
This means that the onset of the CE evolution cannot take place after several thermal pulses as the core of the WD progenitor would grow so that the resulting post-CE WD would be more massive than observed.
In addition, for the initial WD progenitor that we assumed, the onset of CE evolution has to occur no earlier than the beginning of the TP-AGB phase.
If the mass transfer starts when the WD progenitor is in the early-AGB phase, it does not become dynamically unstable \citep[e.g.][]{Ge_2020,Temmink_2023} as the donor star (i.e., the WD progenitor) is not much more massive than its companion (${\approx0.911}$~\Msun), resulting in a WD binary with a much longer orbital period than that observed.
To have dynamically unstable mass transfer with such a low mass ratio, as is required to explain the observed orbital period, the onset of mass transfer has to occur during a \hf, during the first thermal pulses.
This is needed because only during the thermal pulse are the changes in the radius of the TP-AGB star large and do they occur on a very short timescale, which causes mass transfer to be dynamically unstable.

\begin{figure}
\begin{center}
\includegraphics[width=0.99\linewidth]{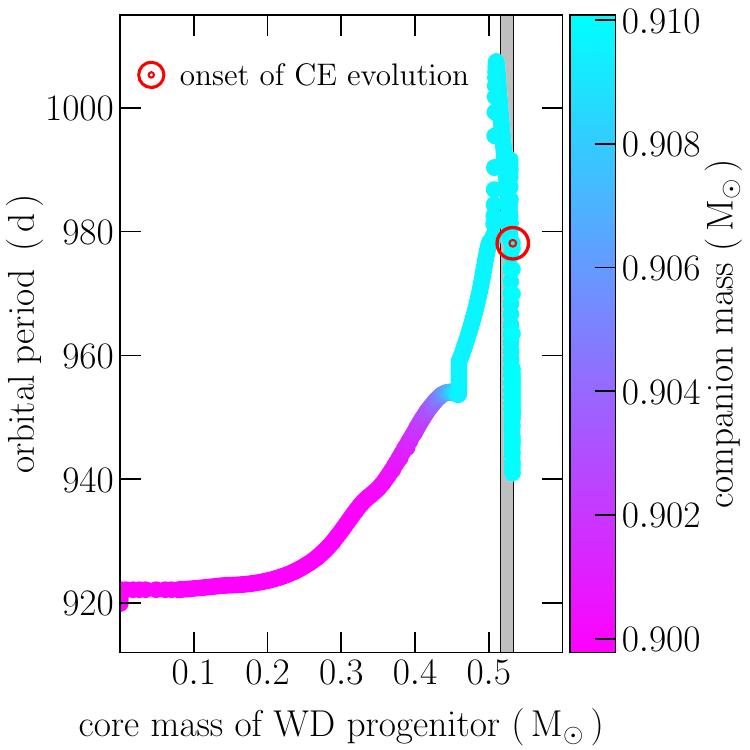}
\end{center}
\caption{Pre-CE evolution of the orbital period with the core mass of the WD progenitor, color-coded with the companion mass. The gray strip in the vertical direction represents the observed WD mass in \koi, while the red solar symbol shows the onset of the CE evolution. When the WD progenitor becomes an evolved first giant branch star, the companion starts to accrete a non-negligible amount of mass. The WD progenitor fills its Roche lobe during the first \hf, as soon as it becomes a TP-AGB star.}
\label{Fig:preCE}
\end{figure}

We provide in Table~\ref{Tab:FormationChannel} an example of a zero-age MS binary that evolves into a binary with properties comparable to those of \koi.
We emphasize that this corresponds to an example because other solutions most likely exist when different values of stellar and binary evolution parameters as well as different initial binary properties are adopted.
The initial masses of the WD progenitor and its companion are $1.5$ and $0.9$~\Msun, respectively.
The orbit is circular with a period of $920$~d.
When the WD progenitor becomes an evolved first giant branch star, wind accretion is no longer negligible and the companion mass slightly increases.
As soon as the WD progenitor becomes a TP-AGB star, it fills its Roche during the first \hf, when its mass is ${\approx1.18}$~\Msun~and its hydrogen-free core has a mass of ${\approx0.53}$~\Msun.
At this moment, the orbital period is ${\approx978}$~d and the companion mass increases to ${\approx0.91}$~\Msun~due to wind accretion.
This pre-CE evolution is shown in Fig.~\ref{Fig:preCE}, in which we include the evolution of the orbital period and the MS companion mass with the core mass of the WD progenitor.

\begin{figure}
\begin{center}
\includegraphics[width=0.99\linewidth]{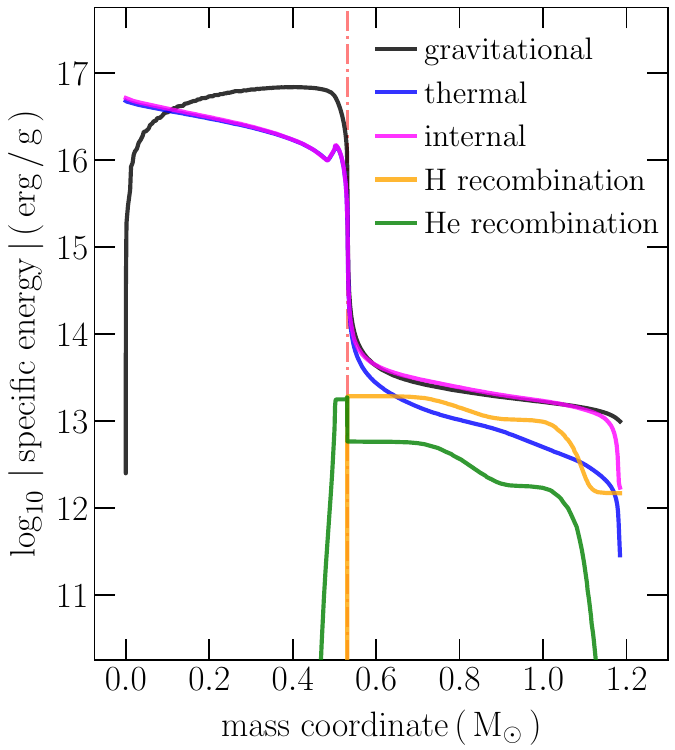}
\end{center}
\caption{Profiles of different specific energies (i.e., gravitational, thermal, thermodynamic internal, and hydrogen and helium recombination) that contribute to the CE binding energy. The vertical line indicates the core mass.}
\label{Fig:CEprofile}
\end{figure}

% --------------------
%  ENERGIES FROM MESA (erg)
% --------------------
% energy-gravitational = -4.900e+46 
% energy-internal = 4.125e+46 
% energy-thermal = 2.426e+46 
% energy-recombination-H = 1.577e+46 
% energy-recombination-He = 3.928e+45 

During CE evolution, different energy sources can contribute to the integrated envelope binding energy.
At the onset of the CE evolution, the gravitational energy is ${\approx-4.9\times10^{46}}$~erg.
The thermodynamic internal energy is ${\approx4.125\times10^{46}}$~erg, with the contribution from thermal energy being ${\approx2.426\times10^{46}}$~erg.
The hydrogen and helium recombination energies are ${\approx1.577\times10^{46}}$ and ${\approx3.928\times10^{45}}$~erg, respectively.
We illustrate in Fig.~\ref{Fig:CEprofile} the profiles at the onset of the CE evolution of these energies.
The CE binding energy we adopted here, which is defined in Eq~\ref{Eq:AlphaINT}, is the gravitational energy plus the thermodynamic internal energy; that is, ${\approx-7.753\times10^{45}}$~erg.

\begin{figure}
\begin{center}
\includegraphics[width=0.99\linewidth]{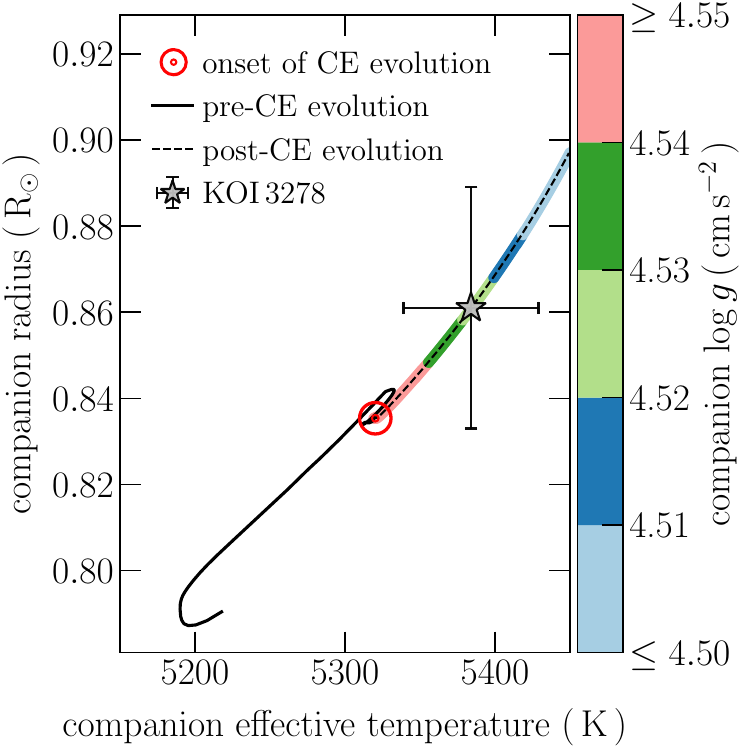}
\end{center}
\caption{Evolution of the MS companion radius with its effective temperature. The black line corresponds to the pre-CE evolution, while the dashed line represents the post-CE evolution. The colors indicate its $\log{}g$ during post-CE evolution. The red solar symbol indicates the onset of the CE evolution and the gray star the observed properties. During pre-CE evolution, the effective temperature and radius increase, but this is reversed due to accretion of a portion of the stellar winds of the WD progenitor. After ${\approx2.2}$~Gyr of post-CE evolution, the MS companion has properties that are similar to those of the G-type star in \koi.}
\label{Fig:HRdiagram}
\end{figure}

The observed orbital period of ${\approx88}$~d can be reproduced if a fraction of ${\approx0.97}$ of the change in the orbital energy is used to unbind the CE.
Afterward, when the post-CE binary is ${\approx2.2}$~Gyr old and the MS companion is ${\approx5.2}$~Gyr old, the properties of the MS star resemble the
observed ones. %like those observed.
We show in Fig.~\ref{Fig:HRdiagram} the pre- and post-CE evolution of the radius of the MS companion as a function of its effective temperature.
For the post-CE evolution, the different colors indicate its $\log{}g$.
During pre-CE evolution, both the effective temperature and the radius of the MS companion initially increase.
This trend is reversed due to the accretion of a fraction of the stellar winds from the WD progenitor for a short period of time.
%
%During post-CE evolution, this trend is reversed back and 
After ${\approx2.2}$~Gyr, the properties of the MS companion are quite similar to those of the G-type star in \koi.
The properties we predict in our modeling with MESA are compared with the observed ones in Table~\ref{Tab:Parameters}.

\begin{table}
\caption{Predicted and observed \citep{Yahalomi_2019} parameters of \koi.}
\label{Tab:Parameters}
\centering
%\small
\setlength\tabcolsep{8pt} % default value: 6pt
\renewcommand{\arraystretch}{1.45} % Default value: 1
\begin{tabular}{lcc}
\hline\hline
%\vspace{-0.2cm}
Parameter & Observed & Predicted \\
%\vspace{0.1cm}
\hline
orbital period (d)                     &  $88.1805$                     &  $88.1805$ \\
WD mass (\Msun)                        &  $0.5250^{+0.0082}_{-0.0089}$  &  $0.531$ \\
\multicolumn{3}{c}{MS companion} \\
mass (\Msun)             &  $0.911^{+0.023}_{-0.026}$     &  $0.910$ \\
radius (\Rsun)           &  $0.861^{+0.028}_{-0.023}$     &  $0.861$ \\
$T_{\rm eff}$ (K)        &  $5384^{+45}_{-44}$            &  $5385$ \\
$\log\,g$ (cm\,s$^{-2}$) &  $4.525^{+0.028}_{-0.035}$     &  $4.527$ \\
age (Gyr)                &  $4.3^{+3.2}_{-2.6}$           &  $5.2$ \\
metallicity              &  $[$Fe/H$]$ = $0.118\pm0.040$  &  $Z=0.024$ \\
\hline
\end{tabular}
\end{table}

%porb = 88.1805
%mwd = 0.531
%m2 = 0.910
%rad = 0.861
%teff = 5385
%logg = 4.527
%t = 5219.1031

We searched for a reasonable formation pathway for \koi~based on a set of assumptions about the parameters of stellar and binary evolution.
%
%We would like to emphasize that the scenario we propose here is most likely independent of these assumptions.
%
By varying the stellar evolution parameters such as the mixing length and the amount of core overshooting, or even by including others not considered here such as semiconvective and thermohaline mixing, gravitational settling, and chemical diffusion, we most likely would be able to reproduce \koi, although different parameters of the zero-age MS binary would be required.
For instance, some of these parameters affect the core mass of the WD progenitor at the onset of the TP-AGB phase, which would then require a zero-age star with a different mass.
Additionally, the TP-AGB phase is a notoriously difficult phase of stellar evolution to model.
It is inherently three-dimensional and out of equilibrium, and many previous works have demonstrated that the physical properties of TP-AGB star models are quite sensitive to the input physics and numerics.
Despite that, we believe the uncertainties in stellar and binary evolution should not affect the validity of our scenario.
A more thorough investigation of mass transfer with TP-AGB donors and the impact of assumptions about the stellar and binary evolution parameters will be presented in another paper.

%%%%%%%%%%%%%%%%%%%%%%%%%%%%%%%%
%%%%%%%%%%%%%%%%%%%%%%%%%%%%%%%%
% NEW SECTION
%%%%%%%%%%%%%%%%%%%%%%%%%%%%%%%%
%%%%%%%%%%%%%%%%%%%%%%%%%%%%%%%%
\section{Discussion}
\label{Discussion}

The formation of close binary stars through CE evolution has been studied with hydrodynamic simulations and has been incorporated in binary population codes such as BSE, SeBa, StarTrack, and binary\_c \citep[e.g.][]{Toonen_2014}.
In order to explain the existence of long-period post-CE 
binaries and close double helium core WDs, both of which require a first mass transfer phase with little or no spiral-in, two approaches have been discussed in the literature.

The first one, largely designed to explain the existence of close double helium core WDs, assumes angular momentum conservation instead of energy conservation \citep{Nelemans_2000}, but the physical interpretation of this approach remains unclear \citep{Webbink_2008,W2012,Ivanova_REVIEW}.
The second idea assumes that recombination energy stored in the envelope, which was used to explain planetary nebula around single stars decades ago \citep{Lucy_1967,Roxburgh_1967,Paczynski_1968}, contributes to the CE ejection process \citep{Han_1994,Han_1995}.  
The question of whether this assumption is reasonable or not has been intensively discussed \citep[e.g.][and references therein]{Soker_2003,Webbink_2008,Ivanova_REVIEW,Zorotovic_2014,Nandez_2015,Ivanova_2015,Ivanova_2018,Soker_2018,Grichener_2018,R2020,K2020,LC2022,GB2022,BelloniSchreiberChapter,Ropke_2023} and observed post-CE binary stars with measured stellar and orbital parameters that seem to require contributions from energies in addition to gravitational energy play a fundamental role in this process \citep{Davis_2010,Zorotovic_2014KOI,M2019,Yamaguchi_2024}.

The first attempt to investigate the formation of \koi~was performed by \citet{Zorotovic_2014KOI}.
These authors reconstructed its evolutionary history and predicted its future using the BSE code, based on the stellar and binary parameters measured by \citet{KruseAgol_2014}.
They found that a small amount of recombination energy, or any other source of extra energy, is required to explain the formation of \koi~through CE evolution; that is, \koi~seemed to provide observational evidence of additional energy playing a role.

%However, throughout the years, little attention was payed to the details of the TP-AGB evolution and its potential impact on the outcome of CE evolution .
%
Over the past few decades, mass transfer from AGB stars leading to CE evolution has been frequently modeled \citep[e.g.][]{Izzard_2018}.
However, the intricate details of TP-AGB evolution have often been ignored, possibly due to the high computational costs involved \citep[e.g.][]{Pols_1998,Dewi_2000,B2008,Davis_2010,Xu_2010,Loveridge_2011,Temmink_2023}.
Nonetheless, there have been some hydrodynamic simulations with TP-AGB donors \citep[e.g.][]{R2020,GB2022}, and TP-AGB evolution is included to some extent in the binary population synthesis code binary\_c \citep{Izzard_2004}.

\citet{Belloni_2024b} have recently suggested that highly evolved TP-AGB stars filling their Roche lobe could be the progenitors of long-period post-CE binaries consisting of massive (oxygen-neon) WDs with AFGK-type MS companions.
This scenario does not work for systems like \koi~because mass transfer is likely to become dynamically stable if the WD progenitor is allowed to lose a significant amount of mass.
Therefore, long-period post-CE binaries with low-mass carbon-oxygen WDs (${\lesssim0.55}$~\Msun) can only originate from unevolved TP-AGB stars.
We here have shown that a solution without contributions from recombination energy also exists for \koi~if the extension and the binding energy of the envelope are consistently calculated during \hfs.
Codes such as BSE and binary\_c are unable to provide the structure of the donor during \hfs, and therefore extra energy during CE evolution is required to reproduce the properties of \koi~if BSE is used (and possibly this remains true for binary\_c).

The results presented in this paper and in \citet{Belloni_2024b} show that the currently available sample of post-CE binaries with accurately measured parameters consisting of a WD and a MS companion star can be explained without contributions from recombination energy. 
An interesting follow-up project would be to 
%incorporate the details of TP-AGB evolution, including~\hfs, into binary population synthesis codes in order to 
compare the predictions of population models with the large recently established samples of WD plus MS star binaries \citep[e.g.][]{nebot-gomez-moranetal11-1,RM2012,Jorissen_2016,Oomen_2018,Escorza_2019,Jorissen_2019,R2021,Shahaf_2023a,Shahaf_2023b,Hallakoun_2023,Brown_2023,Yamaguchi_2024}.

%%%%%%%%%%%%%%%%%%%%%%%%%%%%%%%%
%%%%%%%%%%%%%%%%%%%%%%%%%%%%%%%%
% NEW SECTION
%%%%%%%%%%%%%%%%%%%%%%%%%%%%%%%%
%%%%%%%%%%%%%%%%%%%%%%%%%%%%%%%%
\section{Conclusions}
\label{Conclusion}

We recently showed that long-period post-CE binaries containing massive oxygen-neon WDs can form through CE evolution without assuming extremely efficient 
CE evolution and without assuming that additional energy sources contribute to expelling the envelope. This exercise left \koi~as the one and only observed post-CE binary that has been claimed to provide evidence for contributions from additional energy sources such as recombination energy. 
We here have presented binary evolution simulations with the MESA code 
%and searched for a formation pathway leading to a binary with properties comparable to \koi, which is a long-period system consisting of a low-mass carbon-oxygen WD paired with a G-type main-sequence star.
%
showing that the existence of \koi~can be explained if the WD progenitor fills its Roche lobe during a \hf~occurring with the first thermal pulses on the AGB and that in this case the available gravitational and thermodynamic energy are sufficient to expel the envelope. 
%
%This evolutionary pathway does not require energy sources other than gravitational and thermodynamic internal in the CE energy budget to explain the observed orbital period of \koi.
%

Our results have two fundamentally important implications for understanding CE evolution. First, incorporating the detailed evolution of the early 
and late AGB in reconstructing CE evolution is fundamental to avoid drawing wrong conclusions.  
Second, given our result on \koi, not a single post-CE binary consisting of a WD with a stellar companion provides evidence for additional energy sources such as recombination energy playing a role during CE evolution.  
Nevertheless, \koi~remains an intriguing system. In contrast to the vast majority of observed post-CE binaries, understanding the existence of \koi~requires one to assume that nearly all the available orbital energy is used to unbind the CE. 
%, making \koi~an intriguing system towards a better understanding of CE evolution.

%%%%%%%%%%%%%%%%%%%%%%%%%%%%%%%%
%%%%%%%%%%%%%%%%%%%%%%%%%%%%%%%%
% NEW SECTION
%%%%%%%%%%%%%%%%%%%%%%%%%%%%%%%%
%%%%%%%%%%%%%%%%%%%%%%%%%%%%%%%%
\begin{acknowledgements}

We would like to thank an anonymous referee for the comments and suggestions that helped to improve this manuscript.
We thank the Kavli Institute for Theoretical Physics (KITP) for hosting the program ``WDs as Probes of the Evolution of Planets, Stars, the Milky Way and the Expanding Universe''.
This research was supported in part by the National Science Foundation under Grant No. NSF PHY-1748958.
This research was partially supported by the Munich Institute for Astro-, Particle and BioPhysics (MIAPbP) which is funded by the Deutsche Forschungsgemeinschaft (DFG, German Research Foundation) under Germany's Excellence Strategy -- EXC--2094 -- 390783311.
DB acknowledges financial support from {FONDECYT} grant number {3220167}.
MRS and MZ were supported by {FONDECYT} grant number {1221059}.
MRS was supported by ANID, – Millennium Science Initiative Program – NCN19\_171.

\end{acknowledgements}


\bibliographystyle{aa} % style aa.bst
\bibliography{references} % your references Yourfile.bib

\end{document}